\documentclass[a4paper,11pt]{article}
\pdfoutput=1 

\usepackage{jcappub} 

\usepackage[T1]{fontenc} 

\usepackage{graphicx}
\usepackage{subfigure}
\title{\boldmath Relativistic iron lines in accretion disks: the contribution of higher order images in the strong deflection limit}

\author[a,b]{Giulio Francesco Aldi,}
\author[a,b]{Valerio Bozza}

\affiliation[a]{Dipartimento di Fisica ``E. R. Caianiello'',\\Via Giovanni Paolo Secondo 132, I-84084, Fisciano (SA), Italy}
\affiliation[b]{Istituto Nazionale di Fisica Nucleare, Sezione di Napoli, Italy}

\emailAdd{giuliofrancesco.aldi@sa.infn.it}
\emailAdd{valboz@sa.infn.it}

\abstract{The shapes of relativistic iron lines observed in spectra of candidate black holes carry the signatures of the strong gravitational fields in which the accretion disks lie. These lines result from the sum of the contributions of all images of the disk created by gravitational lensing, with the direct and first-order images largely dominating the overall shapes. Higher order images created by photons tightly winding around the black holes are often neglected in the modeling of these lines, since they require a substantially higher computational effort. With the help of the strong deflection limit, we present the most accurate semi-analytical calculation of these higher order contributions to the iron lines for Schwarzschild black holes. We show that two regimes exist depending on the inclination of the disk with respect to the line of sight. Many useful analytical formulae can be also derived in this framework.}

\begin{document}
\maketitle
\flushbottom
\section{Introduction}
\label{sec:intro}
Black holes reveal their presence in the universe thanks to diffuse material in orbital motion around them that emits electromagnetic radiation particularly in the X-ray band. The emission is attributed to the radiation of inward-spiralling matter in the form of an accretion disk. The dominant features in the X-ray spectrum seen by a distant observer are the iron lines with broad skewed profiles. The energy of the Fe K$\alpha$ line is about 6.4 KeV and the natural bandwidth is a few eV. Relativistic effects and the enormous velocity of the material in the disk are the principal causes of the line deformation \cite{fabian1989} and the typical observed bandwidth are about thousand of KeV \cite{Tanaka1995,Martocchia2002}. Furthermore, it was noted that Fe K$\alpha$ is unaffected by recombination phenomena across the geodesics path near the accretion disk environment. This underlines the reality and the importance of the broad red wing of the line \cite{fabian2008} as a tool to analyze the main characteristics of the central black hole. Fabian et al. constructed a \emph{diskline} model for the relativistic line around a non-rotating black hole \cite{fabian1989}. A model of the relativistic line for a maximally rotating Kerr black hole, based on numerical approach, was developed independently by Kojima \cite{kojima1991}, Laor \cite{laor91} and Arnaud \cite{arnaud1996}. This allowed to extract unique information on the structure, geometry and dynamics of the accretion flow in the immediate vicinity of the central black hole \cite{fabian1989,fabian2008,laor91}. A. Martocchia, V. Karas and G. Matt developed a detailed model of the relativistic effects on both the reflection continuum and the iron line profile \cite{martocchia2000}. M. Dovciak, V.Karas and T. Yaqoob analyzed X-ray spectra of black hole accretion disk by introducing new routines for the XSPEC package \cite{arnaud1996,dovciak2004}. Similarly, Cadez and Calvani developed an advanced code including warped disks and other effects \cite{Cadez2005}. We also mention the fast code developed by Dexter and Agol \cite{DexAgo,DexAgoFra}. In these codes, the photons coming from the relativistic accretion disk are followed along their geodesics in Kerr space using semi-analytical integration in terms of elliptic integrals.

As well known, gravitational lensing is responsible for the creation of multiple images of the same source. In the case of an accretion disk, we will have a direct image from photons coming from the disk side face to the observer, but also a first order image from photons emitted from the opposite side and turning around the black hole. As shown by G. Bao, P. Hadrava and E. \O stgaard, the first order image can contribute to the total flux as much or even more than the direct image \cite{bao1994}. Also it has been shown in Ref. \cite{fuerst} that the most significant contribution to the total flux appears for higher inclination of the accretion disk with respect to the line of sight. Higher order images, generated by photons performing one or more complete revolutions around the black hole, are exponentially suppressed. However, their total contribution can reach a few percents of the total flux and become dominant at their peak frequencies \cite{bao1994}. The problem with higher order images is that they pose the most sever computational challenge for numerical codes. So, apart from the Schwarzschild case, illustrated in detail in Refs. \cite{bao1994,bao1994a,bao1992} using elliptic integrals, higher order images are often neglected. Recently, Johannsen and Psaltis have shown that the higher order images of the accretion disk would build up a bright ring surrounding the shadow, which can be used to estimate the ratio $M/D$ of the black hole \cite{JohPsa}. Perspectives for detecting this feature by VLBI are currently discussed \cite{Joh2012}.

Once we are convinced of the importance of describing higher-order images of disk accretions correctly, in this paper, we propose to apply the well-known approximation technique of the Strong Deflection Limit (SDL), which has been successfully employed in a vast literature to describe the formation of higher order images of isolated sources in a large number of metrics (see e.g. \cite{Bozza2010} and references therein). The Schwarzschild case was considered by Darwin \cite{Darwin1959} as a limit of the exact formula of the deflection angle based on elliptic integrals. The method was then generalized in Ref. \cite{Bozza2002} to any spherically symmetric metrics, and then to rotating black holes and to sources in arbitrary positions \cite{Bozza2003,Bozza2005,Bozza2006,Bozza2007}. Higher orders in the expansion were calculated in Ref. \cite{Iyer2007}. The time delay between different higher order images was derived in Ref. \cite{BozzaMancini2004}. It is interesting to note that alternatives to the Schwarzschild solution arising in several gravitational theories can be typically distinguished by their higher order images pattern, which strictly depends on the derivatives of the metric coefficients, as shown in Ref. \cite{Bozza2002}. In the framework of the study of the iron emission lines created in accretion disks, these analytical approximations can be very useful to set up a simple and accurate treatment of the contribution of the higher order images to the spectral lines. Coupled with a numerical treatment for the direct and first order images, this treatment can provide the missing complement to build up the whole picture of the relativistic effects affecting the formation of such high energy emission lines.

In this paper we will achieve this goal through the following steps. In Section 2 we specify the model for a thin accretion disk used thereafter. Section 3 introduces the strong deflection limit formalism on which our method is based. We then calculate all needed functions fully analytically.  Section 4 contains a thorough discussion of the physical appearance of higher order spectral contributions for various ranges of the parameters, supplying several interesting analytical formulae describing peaks, ranges and other features. We discuss the relation to other works and future perspectives in Section 5.

\section{Accretion disk around a Schwarzschild black hole}

Many observations are consistent with the idea that spectra of several stellar-mass black holes in binary system and super-massive black holes in AGN arise from a hot accretion disk. The first detection of K$\alpha$ iron line occurred in the X-ray spectrum of Cygnus X-1 by EXOSAT Observations \cite{barr}. The smeared profile was attributed to relativistic effects such as gravitational and Doppler frequency shift by Fabian et al \cite{fabian1989}. Iron line profiles are the result of the sum of the intensity contributions arising from the whole accretion disk feeding the black hole, where each source element is characterized by its own orbital velocity and by a local gravitational potential. Being aware that many models of the X-ray emissivity of the disk exist, here we simply assume a Keplerian thin accretion disk spiralling around a Schwarzschild black hole. The methodology we are going to introduce for the description of higher order images is independent of the source model and can be easily transferred to more advanced scenarios.

We introduce the Schwarzschild line element as
\begin{equation}\label{metricS}
\mathrm{d}s^2= \left(1-\frac{1}{r}\right)\mathrm{d}t^2-\left(1-\frac{1}{r}\right)^{-1}\mathrm{d}r^2-r^2\mathrm{d}\vartheta^2-r^2 \sin^2\vartheta\,\mathrm{d}\phi^2.
\end{equation}
where we are using units of the Schwarzschild radius $2MG/c^2=1$. We will often use the short notation $\mu\equiv \cos \vartheta$.

With a disk lying on the equatorial plane $\vartheta=\pi/2$, and assuming axial symmetry, the intrinsic emissivity of the disk is expressed by
\begin{equation}\label{em}
  I_e(r_e,\mu_e)=I_0\, r_e^{-q}\,f(\mu_e)\,\delta(E_{em}-E_0)\;\; \mathrm{W\,\,m^{-2}\,\, Hz^{-1}\,\, sr^{-1}},
\end{equation}
where
\begin{equation}\label{angem}
f(\mu_e)=\begin{cases} (1+2.06\mu_e),\\  \ln(1+ \mu_e^{-1}). \end{cases}
\end{equation}
The form of Eq. (\ref{em}) assumes separability of the radial and angular emissivity. For the radial emissivity we assume a power-law with $q\geq2$, which means that the intrinsic emissivity of the radiation decreases with the distance from the central black hole \cite{fabian1989}. A commonly accepted slope for the radial emissivity is $q=3$ \cite{zycki}. Following Svoboda et al. \cite{Svoboda2009}, for the angular emissivity we consider two possible choices: the linear limb darkening  and the limb brightening laws, expressed in Eqs. \eqref{angem}. Both laws have found convincing justifications in the literature, depending on the details of the physics behind the X-ray emission. Since they behave in the opposite way, it will be important to check how the incidence of higher order images in the total profile depends on the specific choice of the angular emissivity profile. $E_{em}$ and $E_{0}$ are respectively the emitted energy and the rest energy of the line. The inner radius of the disk is limited by the innermost stable circular orbit (ISCO), which for a Schwarzschild black hole is $r_{in}=3$.  The outer radius may vary for each astrophysical case, being determined by the environment surrounding the black hole. In our numerical examples we will set $r_{out}=20$.

The photons emitted by the accretion disk are strongly affected by Doppler shift, due to the high velocities reached by the emitting matter. They also have to climb the gravitational potential well generated by the black hole and are also strongly lensed. In order to describe all these effects, we need to follow the photon geodesics from the emission to the observation.

We consider a generic observer with coordinates $(r_{o}, \vartheta_o, \phi_o)$, where $r_{o}$ is the radial distance between the observer and the black hole, while $\vartheta_o$ is the inclination of the accretion disk w.r.t. the line of sight (we will often use $\mu_o\equiv \cos \vartheta_o$). We choose the system coordinates so as to have $\phi_o=\pi$.

We then define the angular coordinates in the observer sky $(\theta_1, \theta_2)$ in such a way that the black hole is located in $(0,0)$ with its polar axis projected along $\theta_2$; the accretion disk will thus have a line of nodes along the $\theta_1$ axis, where its diameter is exactly orthogonal to the line of sight.

The observed flux is obtained by summing the contributions for each sky element $\mathrm{d}\theta_1\mathrm{d}\theta_2$. In this element we find photons coming from a particular source element $\mathrm{d}r_e\mathrm{d}\phi_e$ with emission angle $\mu_e$. However, the observed intensity is decreased by the cube of the frequency shift factor $g\equiv \nu_o/\nu_e$, in agreement with the Liouville theorem. Summing up, the observed flux reads
\begin{equation}\label{fluxinte}
F_o(\mu_o,E_0)=\int\int g^3 I_e(r_e,\mu_e)\,\mathrm{d}\theta_1\mathrm{d}\theta_2.
\end{equation}

This integral can be done either by scanning the observer sky and following the photons back to their respective source elements, or by transforming the integration domain to the disk plane. We will take the second path and obtain semianalytic results for the contribution of higher order images to this integral.

\section{Gravitational lensing in the strong deflection limit}

\subsection{Photon geodesics in Schwarzschild}

The treatment of strongly deflected photons presented in Ref. \cite{Bozza2002} assumes equatorial motion from the very beginning, exploiting the spherical symmetry of the metric. In the case of an accretion disk, it is convenient to identify the equatorial plane with the disk plane. Then, photons have to move off from the equatorial plane, in general. Of course, since we are still dealing with a Schwarzschild metric, for each photon we can identify a fixed plane of motion, but it is finally convenient to solve the geodesics equations in full generality, including the polar motion. This is also useful in view of future extensions to the Kerr case.

The geodesics equations can be easily integrated with the Hamilton-Jacobi method developed by Carter \cite{Carter1968} and also reported by Ref. \cite{chandra}.
\begin{equation} \label{eqRint}
\pm\int \frac{1}{\sqrt{R}}\, \mathrm{d}r=\pm\int\frac{1}{\sqrt{\Theta}}\,\mathrm{d}\vartheta,\\
\end{equation}
\begin{eqnarray}\label{eqthint}
\phi_f-\phi_i= J \int \frac{\csc^2\vartheta}{\sqrt{\Theta}}\mathrm{d}\vartheta,
\end{eqnarray}
with
\begin{equation} \label{theta}
\Theta=Q -J^2\cot^2\vartheta,
\end{equation}
\begin{equation} \label{R}
R=r^4-(J^2+Q)r^2+(J^2+Q)r.
\end{equation}

These equations are quite popular in their full Kerr version, while here they appear deprived of the terms containing the black hole spin $a$. As stated before, in the present work we concentrate on Schwarzschild metric in order to have clean and simple analytical formulae, leaving spin effects to future works.

The two constants $J$ and $Q$ appearing in Eqs. (\ref{eqRint})-(\ref{R})  are respectively the projection of the specific angular momentum  of the photon along the spin axis and the Carter integral. These can be directly related to the angular coordinates in the sky $(\theta_1, \theta_2)$ from which the photon reaches the observer. We have \cite{chandra}
\begin{equation}\label{theta1}
\theta_1= - \frac{J}{r_{o}\sqrt{1-\mu_o^2}},
\end{equation}
\begin{equation}\label{theta2}
\theta_2=-\sigma r_{o}^{-1}\sqrt{Q-\mu_o^2\frac{J^2}{1-\mu_o^2}},
\end{equation}
where $\sigma=\pm1$ reminds that photons with the same values of $J$ and $Q$ can reach the observer from above or below the black hole. This sign ambiguity will be solved in the next subsection.

The photon trajectories around the black holes are characterized by the existence of an unstable circular orbit. This can be found by imposing
\begin{equation}\label{orbcirc}
R(r_m)=0, \; R'(r_m)=0.
\end{equation}

Using Eq. (\ref{R}), for the Schwarzschild metric we simply get
\begin{equation}
r_m=\frac{3}{2}; \;\; \; (\sqrt{J^2+Q})_m=\frac{3\sqrt{3}}{2}.
\end{equation}
Photons escaping the unstable orbit would be observed at the border of the so-called shadow of the black hole, a circle centered on the black hole with radius
\begin{equation}
\theta_m= \left. \left(\sqrt{\theta_1^2+\theta_2^2}\right)\right|_{J_m,Q_m}=\frac{3\sqrt{3}}{2r_o}.
\end{equation}

This means that photons with constants of motion that are very close to the minimal values will experience very strong deflections, asymptotically approaching the unstable orbit, and appearing very close to the shadow border in the observer's sky. These are the photons that are properly described in the strong deflection limit approach. To this purpose, we introduce a parameter $\xi$ along the shadow border and a parameter $\epsilon$ quantifying the radial distance from the shadow \cite{Bozza2005}. The transformation from $(\theta_1,\theta_2)$ to the variables $(\epsilon,\xi)$ is defined to be
\begin{eqnarray}\label{parameters}
\theta_1(\epsilon,\xi)&=& \theta_{1,m}(\xi)(1+\epsilon), \label{theta1exi}\\
\theta_2(\epsilon,\xi)&=& \theta_{2,m}(\xi)(1+\epsilon), \label{theta2exi}
\end{eqnarray}
where
\begin{eqnarray}
\theta_{1,m} &=&-\frac{3\sqrt{3}}{2 r_o} \xi, \\
\theta_{2,m} &=&-\sigma\frac{3\sqrt{3}}{2 r_o} \sqrt{1-\xi^2}.
\end{eqnarray}
The corresponding values of the constants of motion are
\begin{eqnarray}\label{jmnew2}
J(\xi,\epsilon)&=&J_m(\xi)(1+\epsilon), \\
Q(\xi,\epsilon)&=&Q_m(\xi)(1+2\epsilon),\label{qmnew2}
\end{eqnarray}
with
\begin{eqnarray}
J_m(\xi)&=& \frac{3\sqrt{3}}{2}\xi \sqrt{1-\mu_o^2},\label{Jm} \\
Q_m(\xi)&=&\frac{27}{4}[1-(1-\mu_o^2)\xi^2].\label{Qm}
\end{eqnarray}

The parameter $\xi$ varies in the closed interval $[-1,1]$. We recover prograde photons when $\xi=+1$ and retrograde photons when $\xi=-1$, with respect to our reference polar axis. Photons on polar orbits have $\xi=0$.

The parameter $\epsilon\in[-1,+\infty)$ so that the $(\epsilon,\xi)$ parametrization can cover the whole sky, in general. However, as we shall see shortly, higher order images appear at very small values of $\epsilon$. As a consequence, we will focus on the region $0<\epsilon\ll 1$.

\subsection{The strong deflection limit} \label{sec SDL}

Now, in order to establish the connection between the source element at $(r_e,\phi_e)$ and the coordinates $(\theta_2,\theta_2)$ in the observer sky, we need to solve the integrals in Eqs.  (\ref{eqRint}) and (\ref{eqthint}). The angular ones are trivial in Schwarzschild. They read
\begin{eqnarray}\label{J1}
\pm\int\frac{1}{\sqrt{\Theta}}\,\mathrm{d}\vartheta&=&\frac{2}{3\sqrt{3}} \left[\sigma \arcsin\left(\frac{\mu_o}{\mu_+}\right)+ m\pi \right],\\
\label{j2}
\pm\int\frac{\csc^2\vartheta}{\sqrt{\Theta}}\,\mathrm{d}\vartheta&=& \frac{2}{3\sqrt{3} \xi\sqrt{1-\mu_o^2}} \left[\sigma\arcsin\left(\frac{\mu_o\xi}{\mu_+}\right) + m\pi \right],
\end{eqnarray}
where the amplitude of oscillations in the polar angle is
\begin{equation}
\mu_+\equiv \sqrt{1-\xi^2(1-\mu_o^2)},
\end{equation}
and the exit sign $\sigma$ is
\begin{equation}
\sigma=\pm (-1)^m.
\end{equation}

The integer $m$ represents the number of polar inversions in the photon motion, while the double sign in $\sigma$ is positive for photons emitted upwards from the disk and negative for photons emitted downwards. The exit sign will thus be the same of the initial sign if the number of polar inversions is even and will be opposite if the number of polar inversions is odd. The observer will observe photons above the black hole ($\theta_2>0$) if they are still decreasing their latitude $\mu$ and below the black hole ($\theta_2<0$) if they are still increasing $\mu$. In this way, the sign ambiguity of Eq. (\ref{theta2exi}) is solved.

The radial integral can be expressed in terms of elliptic integrals \cite{Darwin1959,bao1992}. In alternative, we can expand the integrand for small values of $\epsilon$, which is the heart of the strong deflection limit approach to black hole gravitational lensing \cite{Darwin1959,Bozza2002}. Following Ref. \cite{Bozza2007}, we have
\begin{equation}\label{I1sol}
\pm\int \frac{1}{\sqrt{R}}=-\frac{2}{3\sqrt{3}} \log\frac{\epsilon}{\overline{\epsilon} (r_e)},
\end{equation}
with
\begin{equation}
\overline{\epsilon}(r_e)=\frac{216\left(2-\sqrt{3}\right)\left(\sqrt{3r_{e}}-\sqrt{3+r_{e}}\right)} {\sqrt{3r_{e}}+\sqrt{3+r_{e}}}. \label{overlineepsilon}
\end{equation}

With respect to the expressions contained in Ref. \cite{Bozza2007}, we have set $\mu_s=0$ (a source on the equatorial plane), $a=0$ (no
spin for the black hole), and the observer at infinity $r_o\gg 1$.

Putting all terms together and identifying $\phi_i=\phi_e$ and $\phi_f=\phi_o=\pi$, the two geodesics equations (\ref{eqRint})-(\ref{eqthint}) become
\begin{eqnarray}
&& -\log\frac{\epsilon}{\overline{\epsilon}(r_e)} = \sigma \arcsin\left(\frac{\mu_o}{\mu_+}\right)+ m\pi \\
&&\pi-\phi_e= \sigma \arcsin\left(\frac{\mu_o\xi}{\mu_+}\right) + m\pi.
\end{eqnarray}

These equations can be used to express the variables $(\epsilon,\xi)$ as functions of the source element position $(r_e,\phi_e)$. Explicitly, we have
\begin{eqnarray}
&&\xi = -\sigma\frac{\tan \phi_e}{\sqrt{\mu_o^2+\tan^2\phi_e}}  \label{xiphi} \\
&&\epsilon= \overline{\epsilon}(r_e) \hat \epsilon(\phi_e) \label{epsilonrphi}
\end{eqnarray}
with $\overline{\epsilon}(r_e)$ given by Eq. (\ref{overlineepsilon}) and
\begin{equation}
\hat \epsilon(\phi_e)=e^{-m \pi -\sigma \arcsin \sqrt{\mu_o^2\cos^2 \phi_e+\sin^2 \phi_e}}. \label{hatepsilon}
\end{equation}

\begin{figure}
  \centering
  \includegraphics[width=14cm,height=8cm]{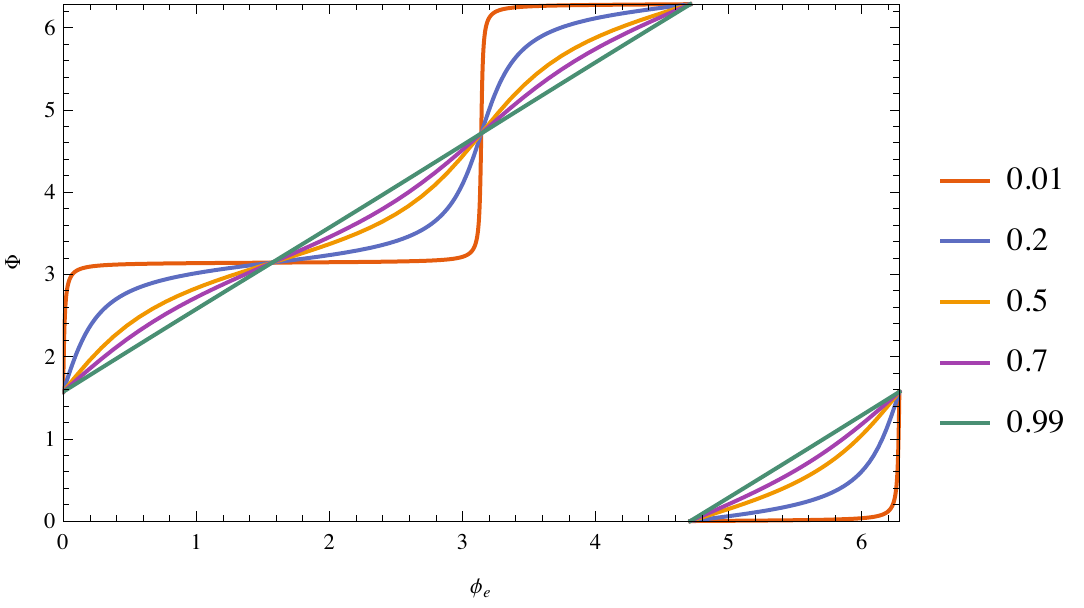}
  \caption{Position angle $\Phi$ in the sky where the image of a source element at azimuth $\phi_e$ is found for a second-order image and for different values of $\mu_o$, from a quasi equatorial observer with $\mu_o=0.01$ to a quasi-polar observer with $\mu_o=0.99$.}\label{azimuth}
\end{figure}

\begin{figure}[htpb]
\centering
  \includegraphics[width=12cm,height=6cm]{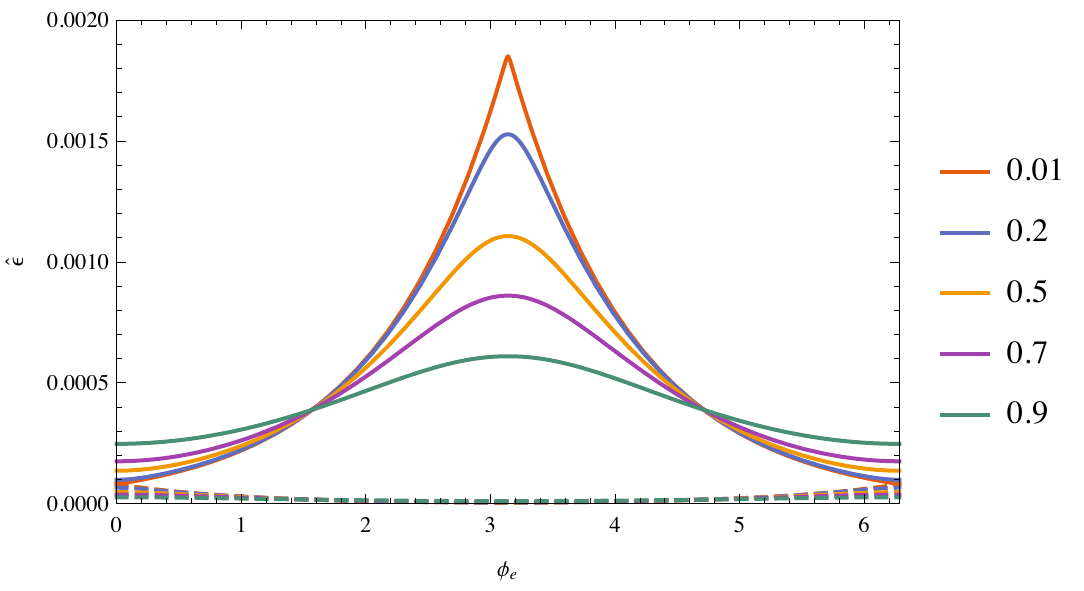}
  \caption{Dependence $\hat \epsilon$ of the displacement of the image from the shadow border as a function of the source azimuth $\phi_e$ for a second-order (continuous line) and third-order image (dashed lines) for different values of $\mu_o$, from a quasi equatorial observer with $\mu_o=0.01$ to a quasi-polar observer with $\mu_o=0.9$.}\label{epshat}
\end{figure}

The first equation reveals that the position angle where a source element appears in the sky only depends on its azimuth $\phi_e$ and the disk inclination
with respect to the line of sight, expressed by $\mu_o$. This is just a consequence of the spherical symmetry of the black hole metric. Fig. \ref{azimuth}
shows the position angle in the observer sky $\Phi$ (given by $-\sigma \arccos (-\xi)$) as a function of the source element azimuth $\phi_e$.
The second equation shows that in the displacement of the image from the shadow the dependencies on $r_e$ and $\phi_e$ factorize. In particular, the displacement varies cyclically with the source azimuth $\phi_e$. The argument of the exponential is mostly dominated by the $-m\pi$, which shows that higher order images appear exponentially closer to the shadow (and are consequently exponentially dimmer). This is illustrated in Fig. \ref{epshat} for images with different order.

\begin{figure}
  \centering
  \includegraphics[width=13cm,height=7cm]{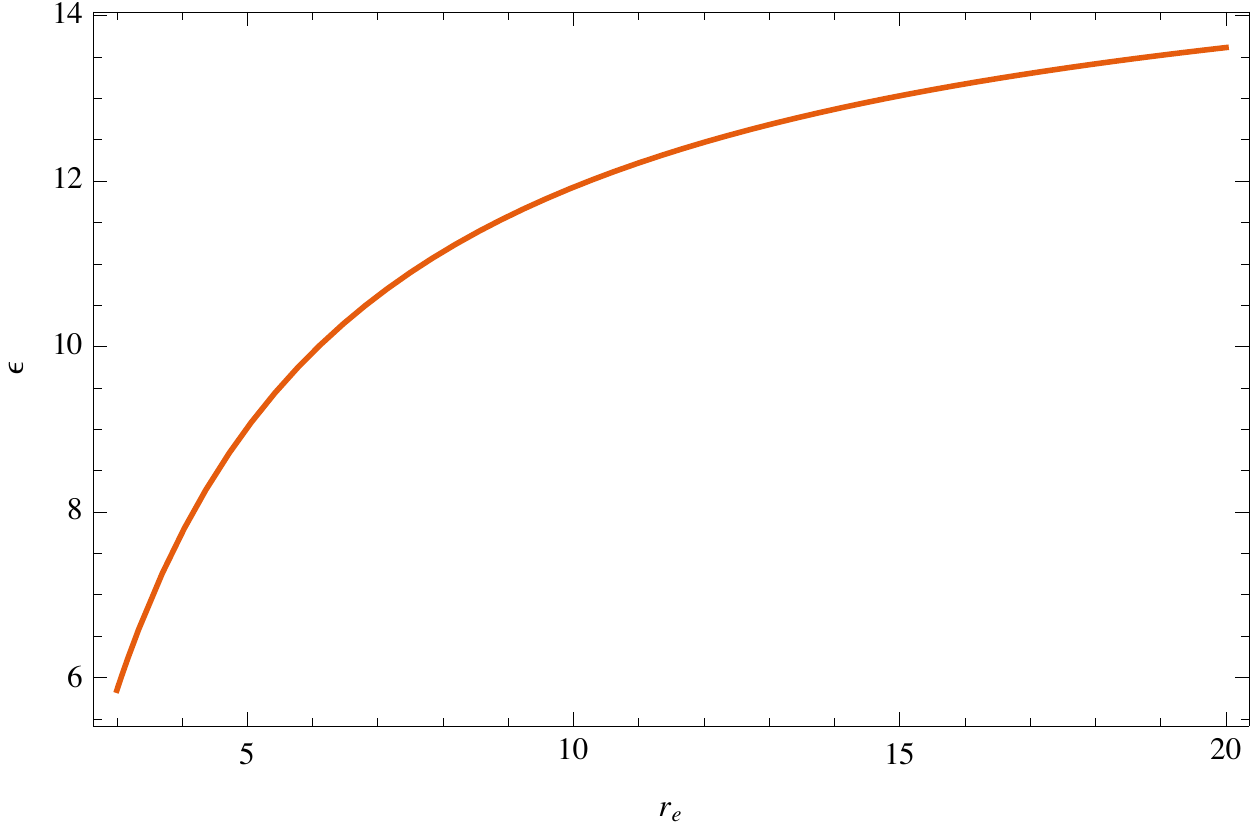}
  \caption{Dependence $\overline{\epsilon}$ of the image displacement from the shadow border on the source radial coordinate $r_e$.}\label{epsbar}
\end{figure}

Let us clarify the definition of the order of an image. The zero-order image is generated by the photons that perform the minimum number of inversions in polar motion. In the case of an accretion disk, we have to distinguish two halves, taking the observer azimuth $\phi_o=\pi$ as the reference. The half of the disk closer to the observer lies at $\pi/2<\phi<3\pi/2$. Photons emitted from this side can reach the observer without any inversion in the polar motion. The zero-order image is thus obtained for $m=0$ and images of order $n$ have $m=n$. Photons emitted from the far side of the disk $-\pi/2<\phi<\pi/2$ necessarily have at least one inversion in the polar motion. Their zero-order image thus has $m=1$ and images of even order $n$, are obtained for $m=n+1$. For odd $n$ we made similar considerations, with the only difference that  $m=n+1$ for the near side of the accretion disk, while $m=n$ for the other. This definition of the order of the images is equivalent to that given in Ref. \cite{bao1994}. We remind that the strong deflection limit is an expansion for small $\epsilon$, which is valid for $m\geq2$. The zero-order image is formed by weakly deflected photons, while the first order one is formed in an intermediate regime. Eq. (\ref{epsilonrphi}) correctly describes images of order higher than one with an accuracy of the order $\epsilon^2$. Finally, the dependence on the radial distance of the source is given by $\overline{\epsilon}(r_e)$ (Eq. (\ref{overlineepsilon})) and shown in Fig. \ref{epsbar}.
Of course, through Eqs. (\ref{theta1exi}) and (\ref{theta2exi}) we can translate the radial displacement $\epsilon$ and the azimuthal parameter $\xi$ in the observer sky coordinates.

\subsection{Magnification}

Once we have recovered the positions of the higher order images of each source element of our accretion disk, we need to calculate the magnification factor. Actually, we have no notion of an unlensed source in the framework we are analyzing. We use the term magnification to refer to the factor relating the image area element $d\theta_1d\theta_2$ to the corresponding source area element $r_edr_e d\phi_e$. This is just the Jacobian of the transformation $(\theta_1,\theta_2) \rightarrow (r_e,\phi_e)$, which is what appears in Eq. (\ref{fluxinte}) when we change the integration variables so as to perform the computation by scanning the accretion disk. Each source element will be mapped to an image element by gravitational lensing. Given the Liouville theorem, the contribution of each element will be more and more relevant the higher the value of this Jacobian.

Using Eqs. (\ref{theta1exi})-(\ref{theta2exi}) and then Eqs. (\ref{xiphi})-(\ref{epsilonrphi}), saving only lowest order terms in $\epsilon$, we get
\begin{eqnarray}
& j_1&= \frac{\partial\theta_1}{\partial \epsilon}\frac{\partial\theta_2}{\partial \xi}-\frac{\partial\theta_1}{\partial \xi}\frac{\partial\theta_2}{\partial \epsilon}= -\sigma \frac{27}{4\sqrt{1-\xi ^2}} \\
& j_2&= \frac{\partial\epsilon}{\partial r_e}\frac{\partial\xi}{\partial \phi_e}-\frac{\partial\epsilon}{\partial \phi_e}\frac{\partial\xi}{\partial r_e} \nonumber \\ &&=-\sigma \frac{648\left(2-\sqrt{3}\right)}{\sqrt{3r_e}\sqrt{3+r_e} \left(\sqrt{3r_e}+\sqrt{3+r_e} \right)^2} \frac{\hat\epsilon(\phi_e)\mu_o^2\cos\phi_e}{\left(\mu_o^2\cos^2 \phi_e+\sin^2 \phi_e\right)^{3/2}}.
\end{eqnarray}

These combine to a magnification
\begin{equation}
A=\frac{j_1 j_2}{r_e} = \frac{4374\left(2-\sqrt{3}\right)}{r_e\sqrt{3r_e}\sqrt{3+r_e} \left(\sqrt{3r_e}+\sqrt{3+r_e} \right)^2} \frac{\hat\epsilon(\phi_e)\mu_o}{\mu_o^2\cos^2 \phi_e+\sin^2 \phi_e} \label{magnification}
\end{equation}

\begin{figure}[htpb]
  \centering
  \includegraphics[width=13.5cm,height=8.5cm]{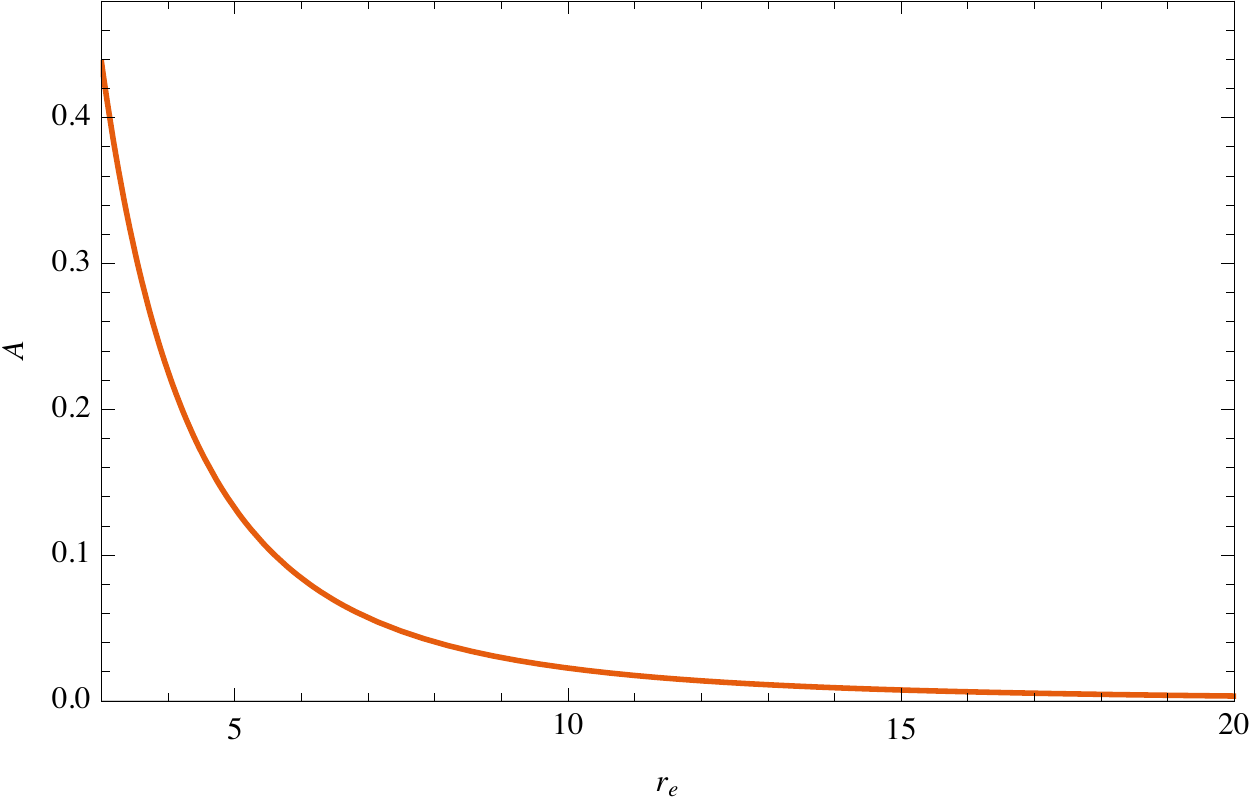}
  \caption{Magnification $A$ in terms of $r_e$ for a generic observer.}\label{Jr}
\end{figure}

\begin{figure}
  \centering
  \includegraphics[width=15cm,height=9cm]{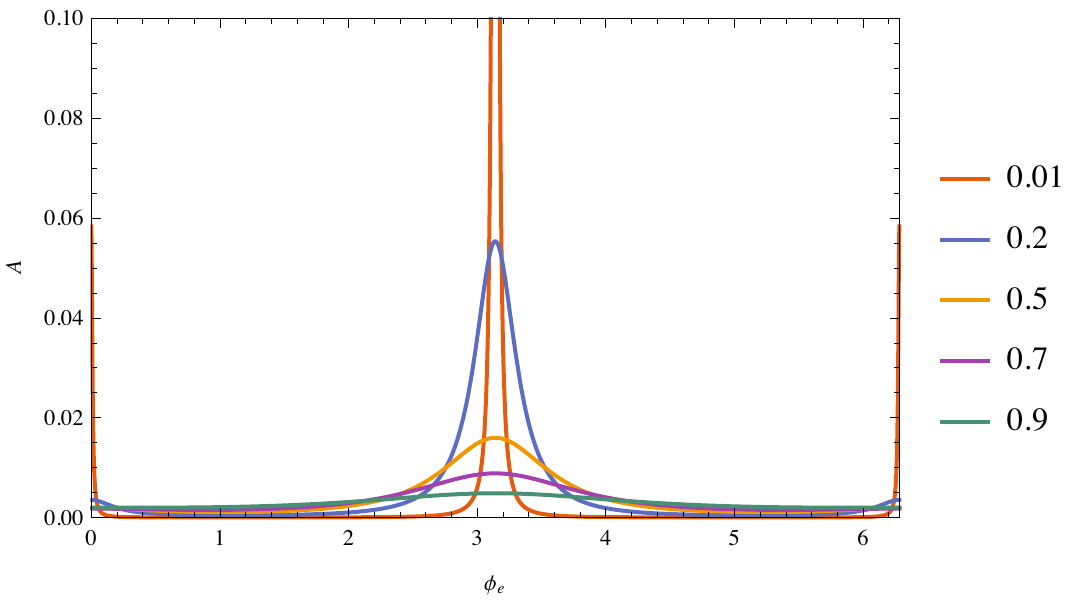}
  \caption{Magnification $A$ for a second-order image in terms of $\phi_e$ for different values of $\mu_o$, from a quasi equatorial observer with $\mu_o=0.01$ to a quasi-polar observer with $\mu_o=0.9$. The two peaks correspond to the caustic points in $\phi_e=0$ and $\phi_e=\pi$. All even order images have the same magnification, though depressed by a $e^{-m\pi}$ factor. Also odd order images have the same structure, though shifted by $\Delta\phi=\pi$. }\label{Jphi}
\end{figure}

The radial and azimuthal dependencies are factorized even in the magnification factor. Fig. \ref{Jr} shows the radial dependency and Fig. \ref{Jphi} the azimuthal dependency for different values of the observer latitude $\mu_o$. Note that the magnification tends to diverge for $\mu_o\rightarrow 0$ in the two caustic points at $\phi_e=0$ and $\phi_e=\pi$. The source elements at these two azimuths will typically give the maximal contribution to the total flux. In particular, even order images have the highest peak at the retro-lensing caustic point $\phi_e=\pi$, corresponding to source elements on the same side of the observer, while odd order images have the highest peak at the standard lensing caustic point $\phi_e=0$, corresponding to source elements on the opposite side.


\subsection{Frequency shift factor} \label{sec redshift}

The photons emitted by each element of the accretion disk undergo Doppler and gravitational redshift. This can be calculated as the ratio of the scalar product between the photon momentum and the observer velocity and the scalar product between the photon momentum and the emitter velocity
\begin{equation}\label{g}
  g=\frac{(g_{\mu\nu}k^\mu u^\nu)_{obs}}{(g_{\mu\nu}k^\mu \dot{x}^\nu)_{em}}.
\end{equation}

If we assume that the velocities of the emitting particles are dominated by the Keplerian rotation, we have
\begin{equation}
\dot{x}^\mu=\left(\sqrt{\frac{2r_e}{2 r_e-3}},0,0,\frac{1}{r_e \sqrt{2 r_e-3}}\right). \label{xmu}
\end{equation}

\begin{figure}
  \centering
  \includegraphics[width=15.5cm,height=15.5cm]{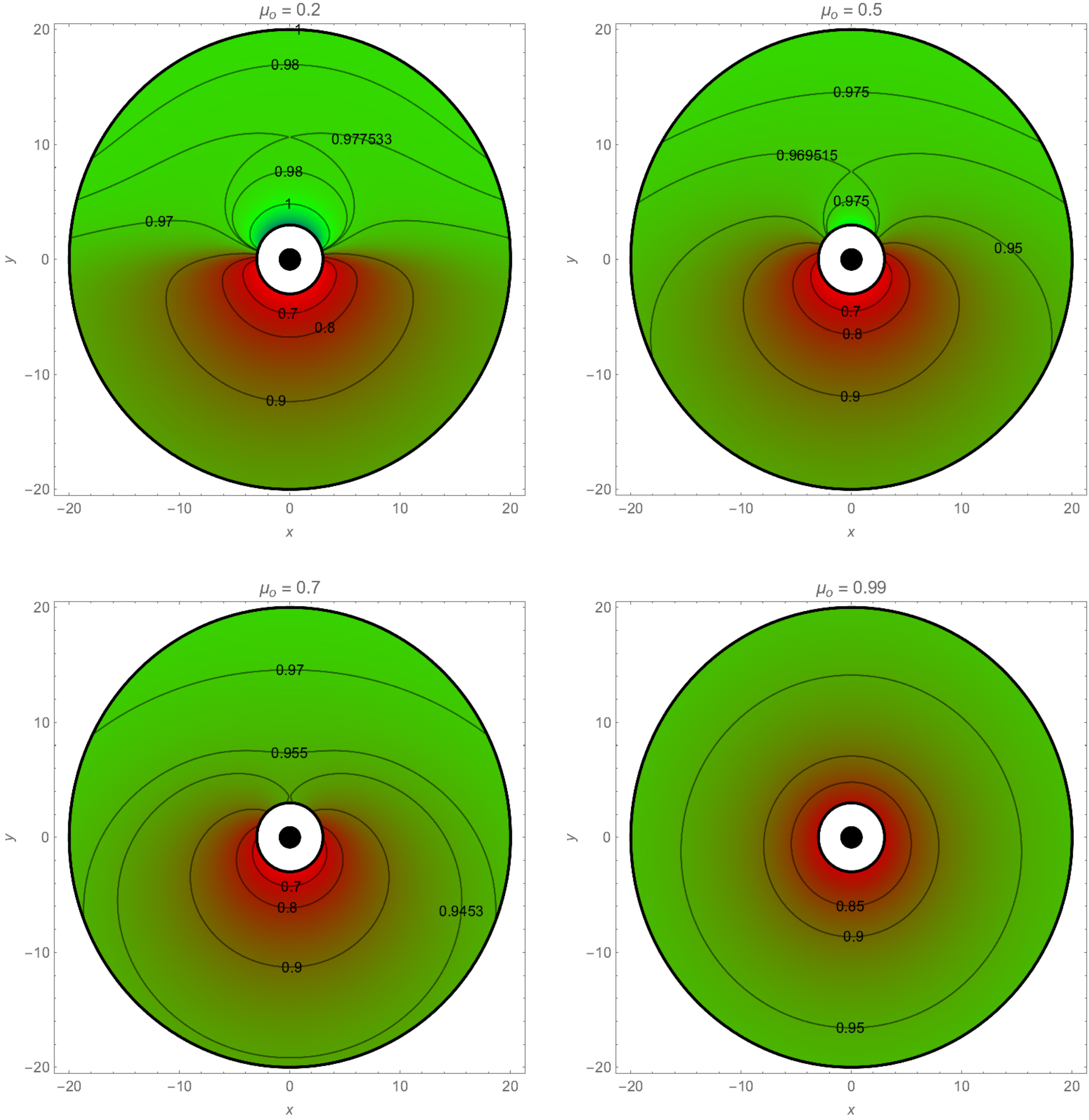}
  \caption{Frequency shift factor contours on the equatorial plane for different values of $\mu_o$ for strongly deflected photons. We remind that the observer is on the left side ($\phi_o=\pi$; the plot is in pseudo-Euclidean coordinates: $x=r \cos \phi,y=r \sin \phi$). The disk is rotating counterclockwise, but photons contributing to higher order images are emitted quasi-radially. Only close enough to the black hole we can find some red(blue)-shift.}\label{gmu}
\end{figure}

\begin{figure}
  \centering
  \includegraphics[width=12.5cm,height=7.5cm]{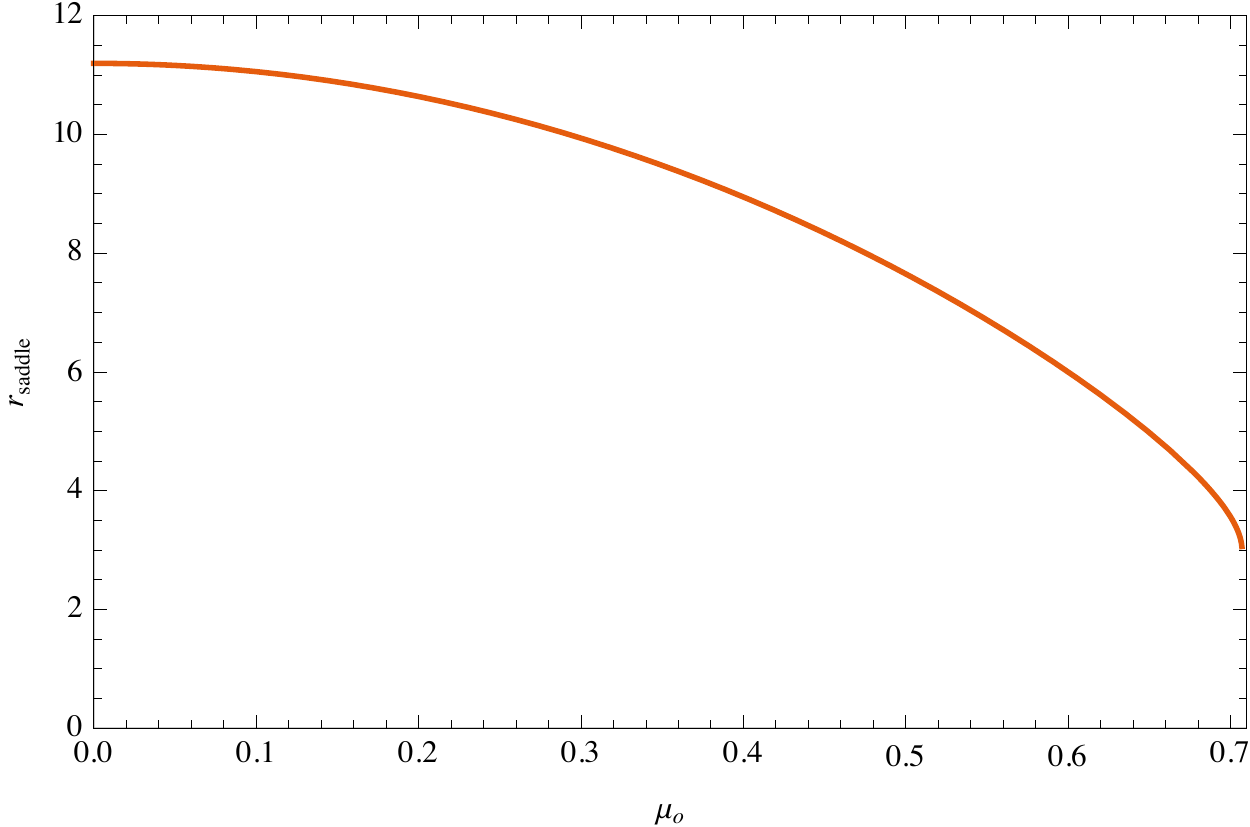}
  \caption{Position of the saddle point in terms of $\mu_o$. }\label{saddle}
\end{figure}

The photon momentum is completely determined  by the constants of motion anywhere along the trajectory:
\begin{equation}
k^\mu=\left(\frac{r}{r-1},-\sqrt{1-\frac{J^2+Q}{r^2} \left(1-\frac{1}{r}\right)},\pm\frac{\sqrt{Q}}{r^2},\frac{J}{r^2}\right). \label{kmu}
\end{equation}
For an observer at rest at infinity $u^\mu\equiv(1,0,0,0)$, the numerator is just $1$. Evaluating (\ref{kmu}) at $r=r_e$ and expressing $J$ and $Q$ in terms of $r_e$ and $\phi_e$ using Eqs. (\ref{jmnew2}), (\ref{qmnew2}), (\ref{xiphi}) and (\ref{epsilonrphi}), we have
\begin{equation}\label{shiftfrequenze}
  g=\frac{ 2r_e \sqrt{2 r_e-3}}{2\sqrt{2}  r_e^{3/2}- 3\sqrt{3}\sqrt{1-\mu_o^2}\xi(\phi_e)}.
\end{equation}
Fig. \ref{gmu} shows the contours of the frequency shift function for different values of $\mu_o$. Since we are considering higher order images, the photons contributing here are those emitted from each disk element in a direction grazing the black hole and then scattered toward the observer.
For any source element, these photons are emitted quasi-radially, thus having very small Doppler shift for emitters in Keplerian rotation. This is reflected by the fact that the denominator of Eq. (\ref{shiftfrequenze}) is typically dominated by the first term, for large $r_e$, independently of $\phi_e$, the second term being subdominant. So, at large enough radii, the contours become circular. However, close enough to the black hole, we have an asymmetry between photons emitted from the side at  $\phi=+\pi/2$, which benefit from some blueshift and photons emitted from the $\phi=-\pi/2$ side, which are further redshifted. So, while on the latter side, the contours stay further from the black hole and approach it monotonically without surprises, on the $\phi=+\pi/2$ we may have an inversion in redshift, with the Doppler term dominating at some point. This determines the existence of a saddle point in redshift on the $\phi=\pi/2$ side, which is visible in the contours shown in Fig. \ref{gmu} for an observer with $\mu_o=0.2, 0.5$ and $0.7$. In order to calculate the position of this saddle point, we impose
\begin{eqnarray}
\frac{1}{r_e}\frac{\partial g}{\partial \phi_e}&=&0 \\
\frac{\partial g}{\partial r_e}&=&0.
\end{eqnarray}
The first equation is only solved for $\phi=\pm\pi/2$. The second equation becomes
\begin{equation}\label{numzero}
 \sqrt{2} r_e^{3/2}-3\sqrt{3} \sqrt{1-\mu_o^2}\, (r_e-1)=0.
\end{equation}
This cubic equation has one root inside the event horizon for $\mu_o>1/\sqrt{2}$ and three roots for $\mu_o<1/\sqrt{2}$. In this regime, we only find one root with $r_e>r_{in}=3$, which lies on the disk. The existence of this saddle point for low enough inclinations has important consequences on the peak of the emission line, as we shall see later. In Fig. \ref{saddle} we show the position of the saddle point as a function of $\mu_o$.

\subsection{Emission angle}

Emissivity laws of accretion disks typically have some dependence on the emission angle. In our study, we consider the two cases shown in Eq. (\ref{angem}) with linear limb darkening or limb brightening alternatives. The first one is the classical linear approximation for the limb darkening and is widely used in literature  \cite{laor91,dovciak2004,beck}, while the second one was proposed by Haardt \cite{Haardt1993} in order to model external sources of X-ray photons \cite{Svoboda2009} reflected by the disk. Of course, the right choice of emissivity law depends on the specific origin of the X-ray sources. However, the choice of intrinsic emissivity does not affect our treatment, and it is possible to change the emissivity law in Eq. \eqref{fluxinte} without restrictions.

Whatever the chosen law, we need to calculate the cosine of the emission angle $\mu_e$. This can be determined by the scalar product of the photon momentum $k^\mu$ at the time of the emission and the normal to the disk
\begin{equation}
 n^\mu=(0,0,\frac{1}{\sqrt{-g_{22}}},0)=\left(0,0,\frac{1}{r},0\right).
\end{equation}
We thus have by definition
\begin{equation}\label{mue}
\mu_e= \frac{g_{\mu\nu}k^\mu n^\nu}{g_{\mu\nu}k^\mu \dot{x}^\nu},
\end{equation}
where $\dot{x}^\nu$ is the four-velocity of the source (\ref{xmu}). This gives
\begin{equation} \label{angolodiemissione}
\mu_e=\frac{3\sqrt{3}\sqrt{2 r_e-3} \sqrt{1-(1-\mu_o^2)\xi^2}}{2\sqrt{2} r_e^{3/2}-3\sqrt{3} \xi\sqrt{1-\mu_o^2} }.
\end{equation}
Note that at large distances we have $\mu_e\rightarrow 0$ as photons of our interest are shot in a direction very close to the black hole grazing the disk surface.

It is worth to comment aberration phenomena caused by gravitational and Doppler frequency shift. On the part of the disk included in $0<\phi<\pi$, where the disk is moving towards the observer, the photons are blue-shifted and the emission angle  (\ref{mue}) increases, while the other part of the disk $\pi<\phi<2\pi$ is moving away from the observer and the photons are red-shifted. As a consequences the emission angle decreases. These effects are less significant for higher distances of the source from the black hole.

\section{Numerical integration}

The profile of the emission lines generated in a relativistic accretion disk is obtained by summing the contributions of the image elements seen by the observer in the sky. As anticipated before, we prefer to convert the integral in Eq. (\ref{fluxinte}) to a sum over the surface elements of the disk. The transformation involves the magnification factor (the Jacobian in other words) calculated in Eq. (\ref{magnification}):

\begin{equation}\label{nuovointegrale}
 F_o(\mu_o,E_0)= \int^{r_{out}}_{r_{in}}\int^{2\pi}_0  g^3(r,\phi) \,r^{-3}f(\mu_e)\, A(r,\phi)\, r\,\mathrm{d}r\mathrm{d}\phi,
\end{equation}
where the inner boundary of the disk is at the ISCO $r_{in}=3$, and we assume an outer boundary at $r_{out}=20$, which is a quite conventional value in the literature allowing an easy comparison with the results reported in other papers. From now on, we drop the subscript $e$ to distinguish the emission point.

The integrand in Eq. (\ref{nuovointegrale}) is essentially composed by two main factors: the intrinsic emissivity and the lensing phenomena, as discussed previously. The amplification and the redshift functions are too complicated to allow an analytic resolution of this integral. Therefore, we resort to a simple numerical integration, obtained by tiling the disk with surface elements $r\mathrm{d}r\mathrm{d}\phi$ at which we evaluate the integrand and the redshift function. We can then assign the element contribution to the correct redshift bin.

Our analytical formulation allows us to calculate the boundaries of the redshift interval covered by the emission line. In fact, as already noted, at each fixed radius $r$, the redshift function has a maximum at $\phi=\pi/2$ and a minimum at $\phi=-\pi/2$. Equivalently, we can obtain the maximum (minimum) value by setting $\xi=+1$ ($-1$) in Eq. (\ref{shiftfrequenze}). As the disk extends from $r_{in}$ to $r_{out}$, we just have to evaluate the redshift function at these radii and read the highest and lowest values obtained. The possible existence of a saddle point in the disk does not represent a problem. In fact, as seen in Section \ref{sec redshift}, this point lies at $(r=r_{saddle},\phi=\pi/2)$. It will be certainly lower than $g(r_{in},\pi/2)$ and $g(r_{out},\pi/2)$ since it is a minimum in the radial direction. It is also higher than $g(r_{saddle},-\pi/2)$ and then higher than $g(r_{in},-\pi/2)$. Then, it is in the middle of our range defined by $g(r_{in},-\pi/2)$ and $\mathrm{max}[g(r_{in},\pi/2),g(r_{out},\pi/2)]$.

In order to get smooth enough profiles, we use $10^6$ tiles in our numerical integrations. The next section presents the results obtained with this procedure.

Finally, in order to quantify the relevance of higher order images compared to the zero and first order ones, which dominate the line profiles, we also had to repeat the same numerical integration for them. For these images, which are outside the strong deflection regime, the only change is that we use the exact formulae for radial integrals involving elliptic integrals and perform derivatives numerically. Of course, we have taken care of the fact that $\epsilon$ is no longer small and may take arbitrary values.

\section{Results: the relativistic emission lines}

Our framework allows a quasi-analytic description of the contributions of higher order images, which are poorly understood and more difficult to analyze by standard numerical codes. First we note that there is no substantial difference between the shapes of the contributions of images of order $n\geq2$. In fact, the order only enters our formulae through a global factor in the magnification (\ref{magnification}) coming from $\hat\epsilon$ (see Eq. (\ref{hatepsilon})). In practice, each order is suppressed by a factor $e^{-\pi}=0.043$ with respect to the previous one, without any further effects on the line shape. The fact that the magnification factor is shifted by $\Delta \phi=\pi$ in images of different parity does not produce any net effect, thanks to the symmetries of the other functions of $\phi$ (red-shift and angular emissivity).

\begin{figure}
  \centering
  \includegraphics[width=15cm,height=9cm]{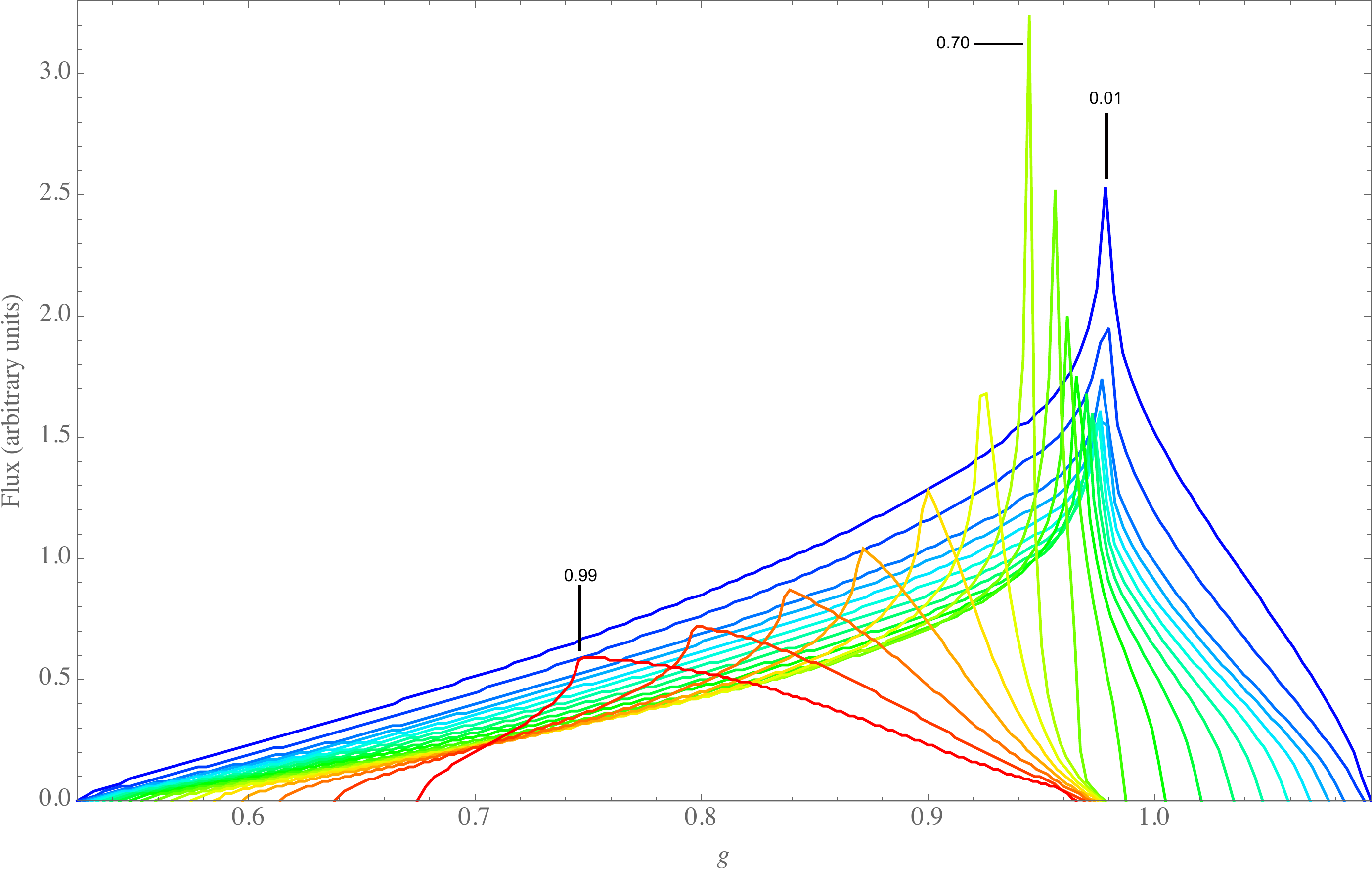}
  \caption{Contribution of higher order images of the accretion disk to a relativistic emission line for observer inclinations ranging from $\mu_o=0.01$ (edge-on) to $\mu_o=0.99$ (face-on) with a step of 0.05. Here the limb darkening law $f(\mu_e)=1+2.06\mu_e$ is adopted.}\label{spettri}
\end{figure}

\begin{figure}
  \centering
  \includegraphics[width=15cm,height=9cm]{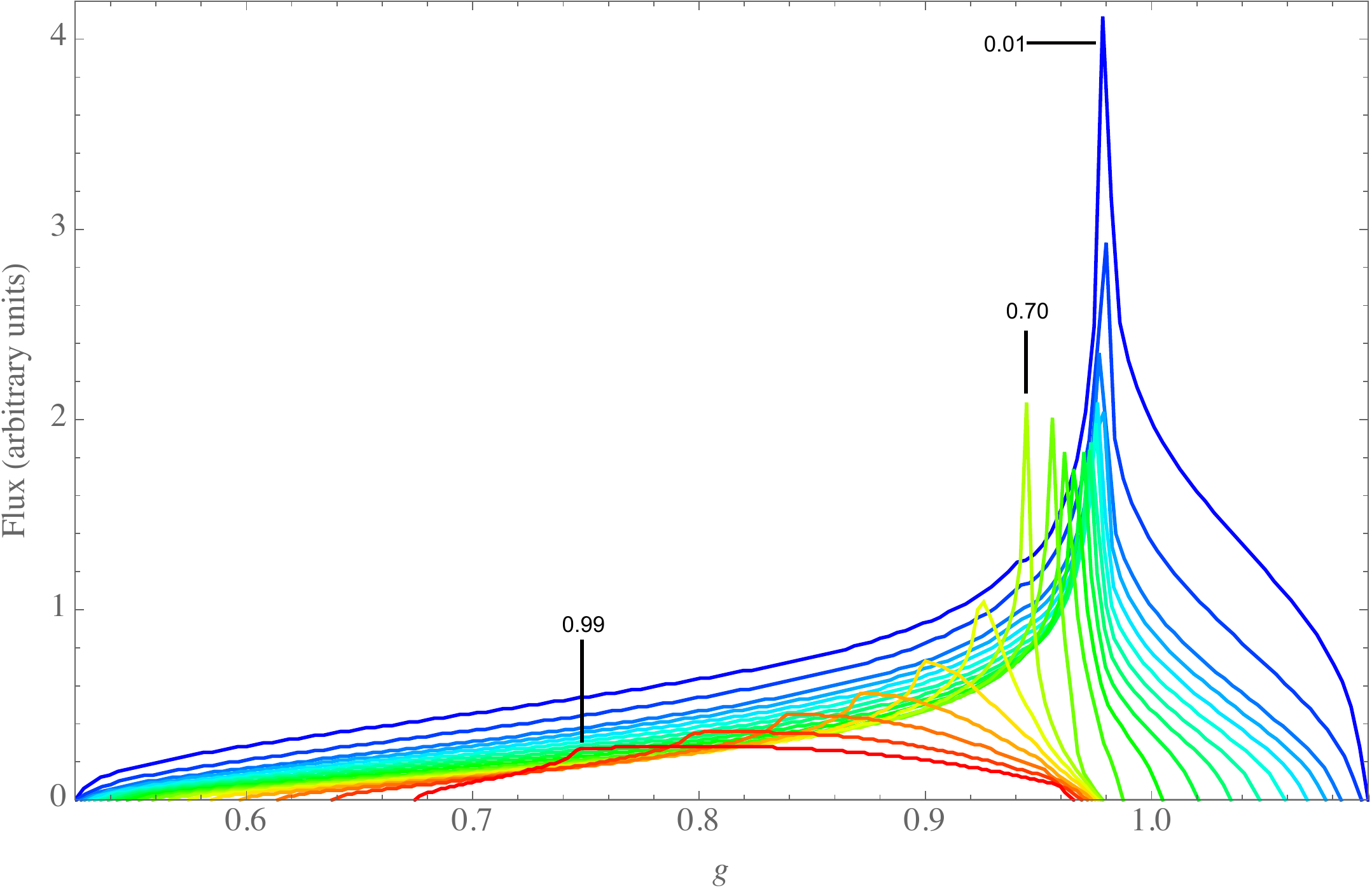}
  \caption{Contribution of higher order images of the accretion disk to a relativistic emission line for observer inclinations ranging from $\mu_o=0.01$ (edge-on) to $\mu_o=0.99$ (face-on) with a step of 0.05. Here we have the limb brightening case $f(\mu_e)=\ln(1+\mu_e^{-1})$.}\label{spettribright}
\end{figure}

Fig. \ref{spettri} and  \ref{spettribright} show the shapes of the generic higher order contribution to the relativistic emission lines produced by an accretion disk around the Schwarzschild black hole for different values of the disk inclination, from $\mu_o=0.01$ (edge-on) to $\mu_o=0.99$ (face-on), with the limb darkening and limb brightening angular emission law  respectively.

The shape of the contribution by higher order images is relatively simple with respect to those by the direct and first order images \cite{bao1992}.
There is only one peak with a broad red wing and a small blue wing. In general, the peak moves toward the red wing when the observer inclination changes from edge-on to face-on. However, there are two distinct regimes that we can spot in Fig. \ref{spettri} and \ref{spettribright}, roughly separated by the $\mu_o=0.7$ curve.
Using the analytical formulae derived in the previous section, we can track these characteristics to their physical origin. In particular, the existence of a saddle point in the redshift function $g(r,\phi)$ is of crucial importance. In fact, there are many area elements of the disk with redshift similar to that of this stationary point and contributing to the neighborhood of its value. Therefore, the position of the peak in the emission line can be very accurately tracked by the redshift of this saddle point.

\begin{figure}
  \centering
  \includegraphics[width=14.5cm,height=9cm]{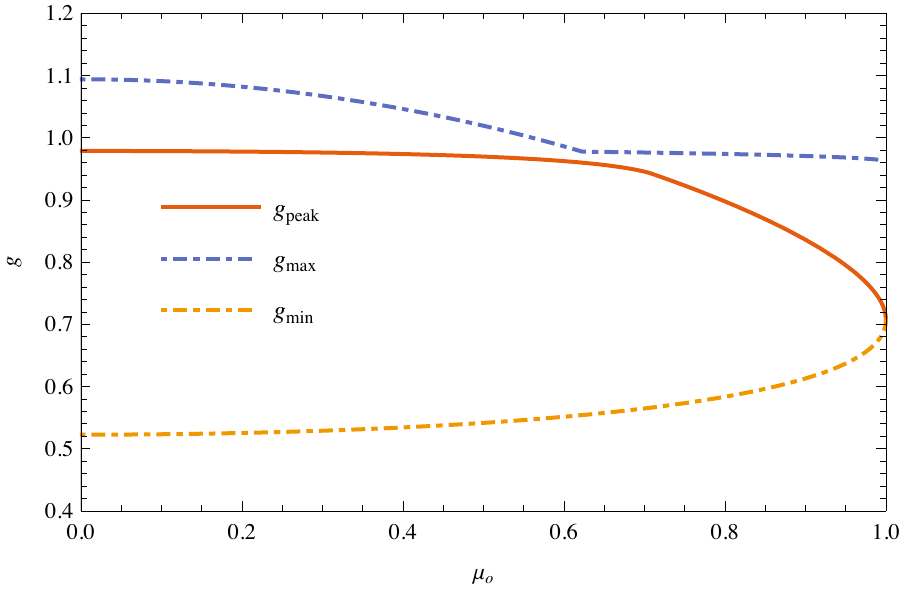}
  \caption{Peak frequency and boundaries of $g$ as functions of $\mu_o$.}\label{gmu0}
\end{figure}

As anticipated in Sec. \ref{sec redshift}, the position of this saddle point is $(r_{saddle},\phi=\pi/2)$, with $r_{saddle}$ given by the largest solution of the cubic equation (\ref{numzero}). Evaluating the redshift (\ref{shiftfrequenze}) at this point, we obtain the peak position analytically. The saddle point exists only for $\mu_o\leq 1/\sqrt{2}=0.707$. At this critical value, the saddle point reaches the ISCO and then disappears.
For $\mu_o > 1/\sqrt{2}$, the disk becomes more equally covered by all redshifts in the interval. The peak then occurs at the redshift corresponding to the blue-shifted side of the ISCO, where the highest emissivity is found. Analytically, this is obtained setting $r=3$ and $\phi=\pi/2$ in Eq. (\ref{shiftfrequenze}):
\begin{equation}
g_{peak_{\mu_o>1/\sqrt{2}}} = \frac{2}{2\sqrt{2}-\sqrt{1-\mu_o^2}}.
\end{equation}
Fig. \ref{gmu0} shows the peak frequency along with the boundaries of the emission line as functions of the observer inclination $\mu_o$. From this picture it is possible to find out the complete structure of the single line starting from the inclination of the accretion disk w.r.t. the line of sight. Also, we reach $g_{max}$ on the ISCO up to $\mu_o\approx0.6$ because of the blueshift doppler, while for greater inclinations it is assumed on  $r_{out}$.
When we consider a limb darkening angular emission law, as the saddle point moves to the inner edge of the accretion disk, the peak frequency benefits from the higher disk emissivity and becomes higher and higher. This is clearly seen in Fig. \ref{spettri}  where the $\mu_o=0.7$ spectrum boasts the highest peak. These two regimes are still present, though less evident, for a limb brightening angular emission law where the emission angle increases as the inclination of the accretion disk increases for a given radius. As a result we obtain lower peaks for higher inclinations, as we can see in Fig. \ref{spettribright}.

\begin{figure}[ht]
  \centering
  \includegraphics[width=14.5cm,height=9cm]{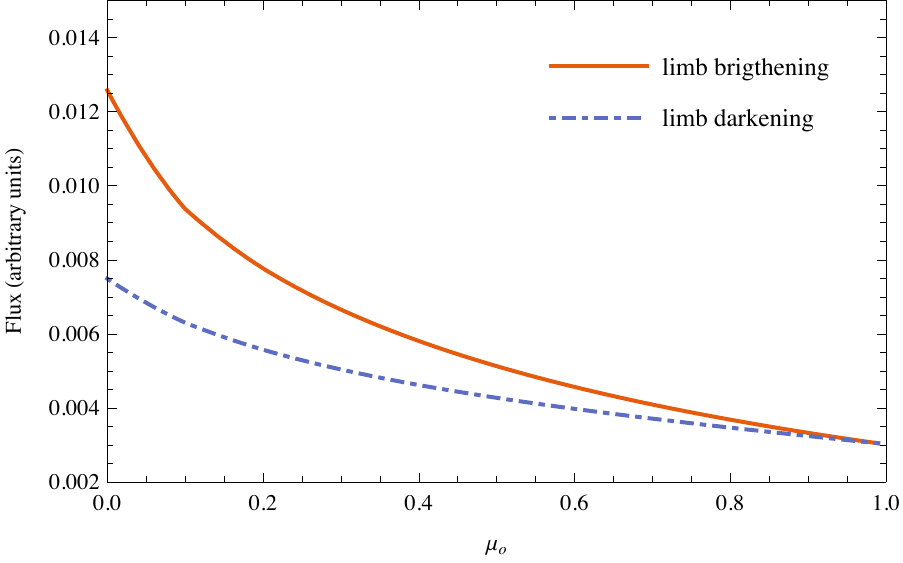}
  \caption{Total flux of each profile in terms of $\mu_o$. dashed line corresponds to the flux evaluated with a limb brightening law, while the other corresponds to the flux evaluated with a limb brightening law.}\label{totalflux}
\end{figure}

Now, let us investigate the effect of lensing magnification. Eq (\ref{magnification}) shows that the regions of the accretion disk behind the black hole ($\phi=0$) and in front of it ($\phi=\pi/2$) get the highest magnification, scaling as $\mu_o^{-1}$. As these two lines contain a representative collection of all redshifts in the line, we expect that the whole profile should be magnified at small $\mu_o$ values. This is indeed what we see in Fig. \ref{spettri} for $\mu_o\lesssim 0.4$: not only the peak becomes higher, but the whole profile is enhanced.
This brings us to the final consideration about the total flux in the line. By summing all bins in our spectra, we get the plot in Fig. \ref{totalflux}, where the flux is maximum for $\mu_o\rightarrow 0$, where the caustic lines asymptotically approach our thin accretion disk. This increase in flux is enhanced in the limb-brightening case, as can be easily expected in an edge-on geometry.

\subsection{Higher vs lower order images}

We have given a semi-analytical description of the contribution of higher order images to the line profiles. However, as already stated previously, this contribution is very small compared to that by the zero and first order images, which dominate the line profile.

\begin{figure}[htpb]
\centering
  \subfigure[\protect\label{fig:fe001}]{ \includegraphics[width=7cm,height=5cm]{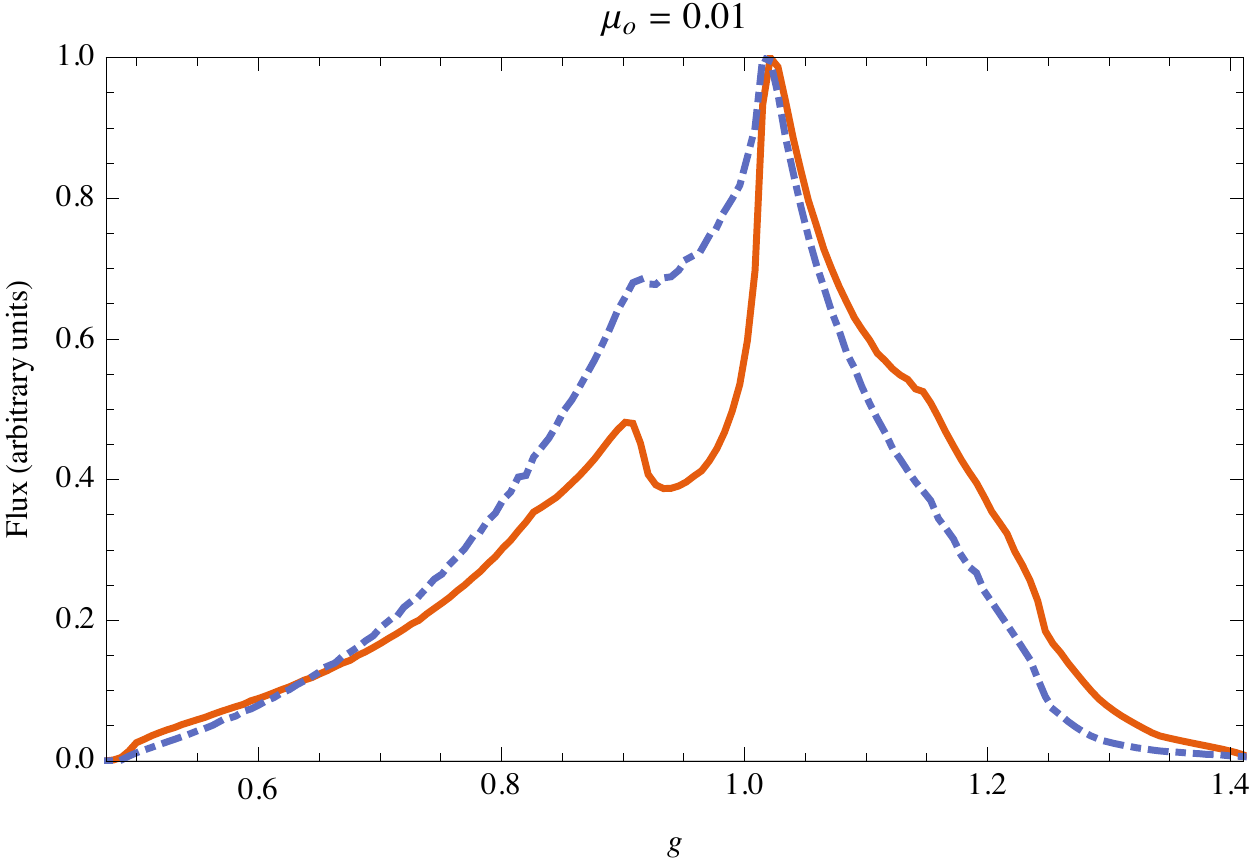}}
  \subfigure[\protect\label{fig:fe070}]{ \includegraphics[width=7cm,height=5cm]{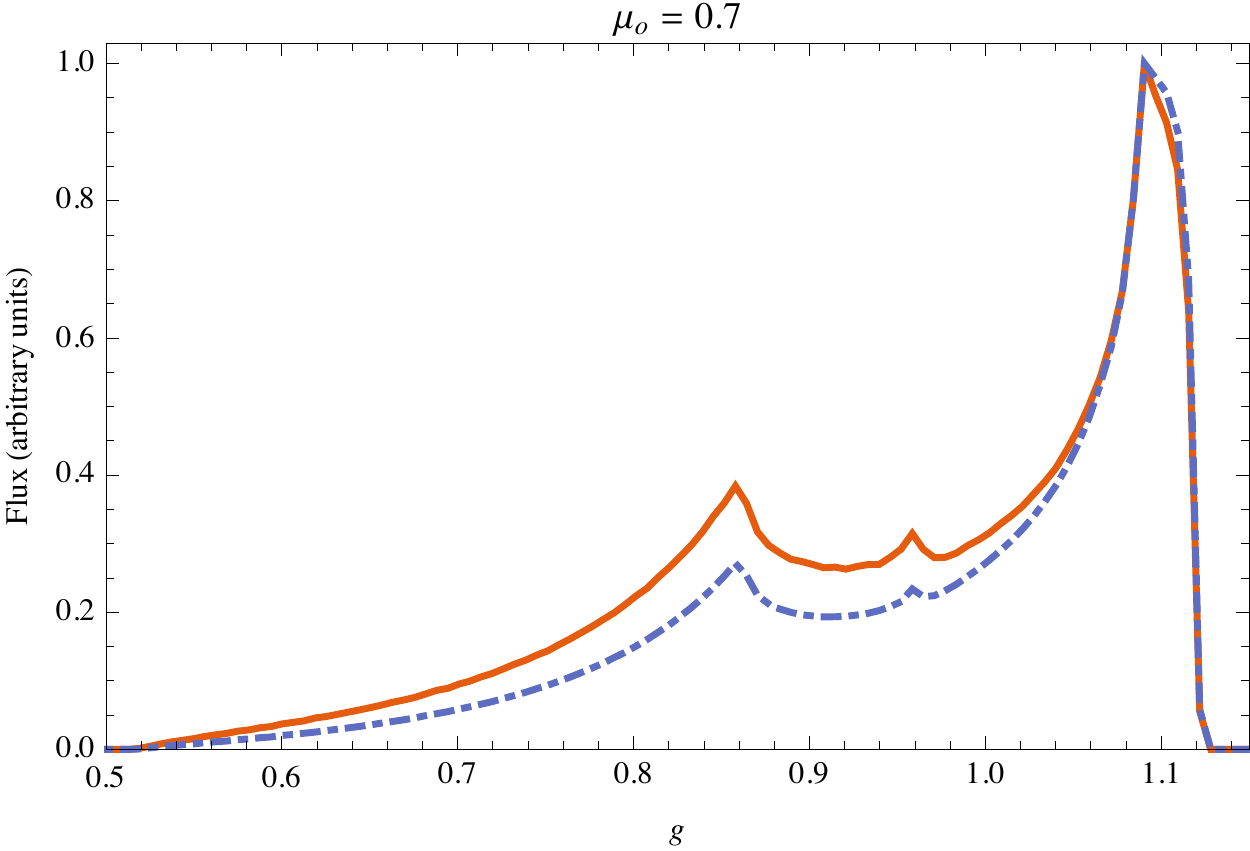}}
  \subfigure[\protect\label{fig:fe099}]{ \includegraphics[width=7cm,height=5cm]{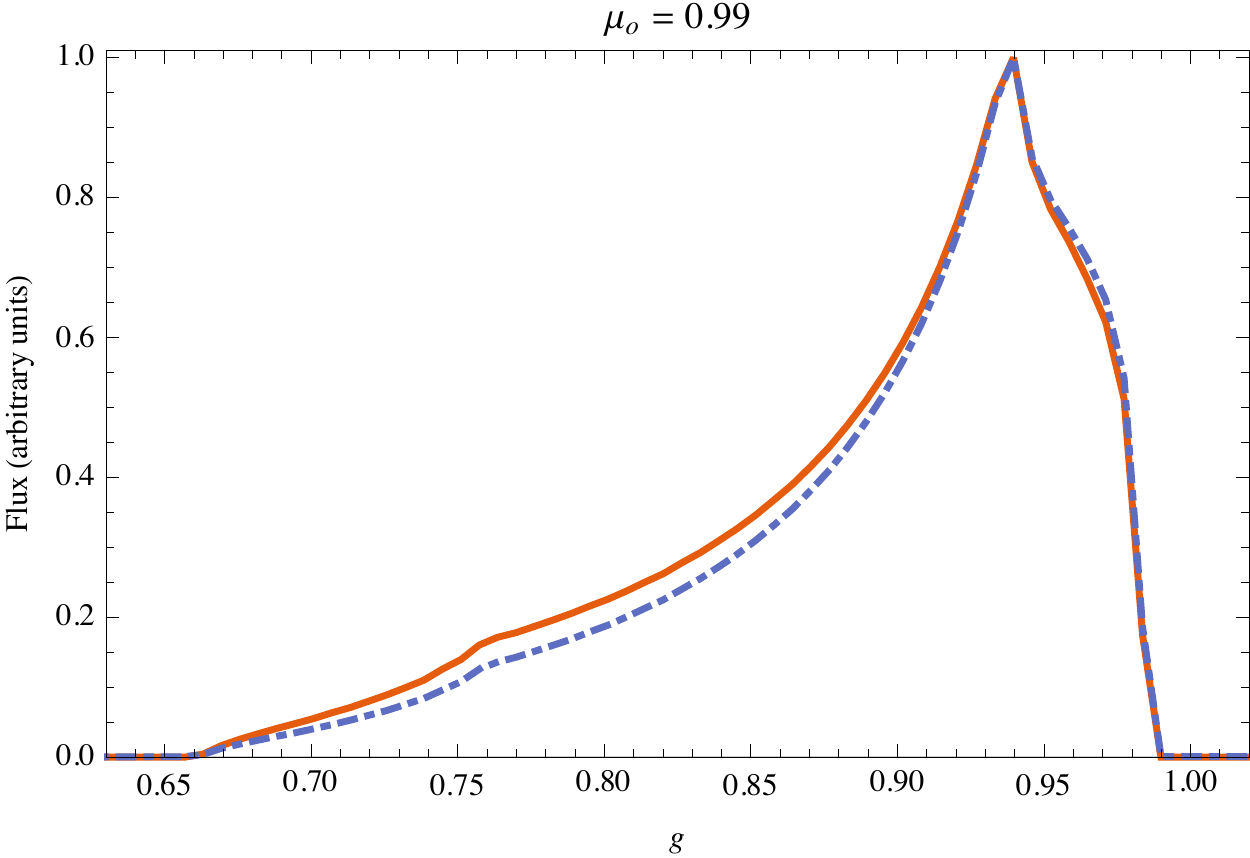}}
  \caption{Iron K$\alpha$ lines for weak deflected photons of zero and first order images. (a) K$\alpha$ line for an edge-on accretion disk $(\mu_o=0.01)$. (b) K$\alpha$ line for an accretion disk with $\mu_o=0.7$. (c) K$\alpha$ line for a face-on accretion disk $(\mu_o=0.99)$. Dashed lines are determined with the limb-darkening law, while continuous lines with the limb-brightening law.}\label{spettriprimi}
\end{figure}

In Fig. \ref{spettriprimi} we show some line profiles as determined by the lower order images only, neglecting higher order images. In particular, we have the cases (a)  $\mu_o = 0.01$ (edge-on accretion disk), (b) $\mu_o = 0.7$  and (c) $\mu_o = 0.99$  (face-on accretion disks). The solid lines are for limb-brightening profile and the dashed lines for limb-darkening profiles. Similar plots can be found in a vast literature \cite{bao1994, Svoboda2009, Svoboda2010}. 

In general, these profiles show two peaks corresponding to blue-shifted and red-shifted photons as emitted by the two sides of the rotating disk. Only in the face-on case, we have a single peak, since no photons benefit from Doppler blue-shift, while all of them are gravitationally red-shifted. In Fig. \ref{fig:fe070} we can also see a third small central peak due to the first order image. The largest difference between limb-darkening and limb-brightening emissivity laws arises in the edge-on case (Fig. \ref{fig:fe001}). Here the minimum between the two peaks becomes shallower if we have a limb-darkening law.

\begin{figure}[ht]
\centering
  \subfigure[\protect\label{fig:ratio1}]{ \includegraphics[width=7cm,height=5cm]{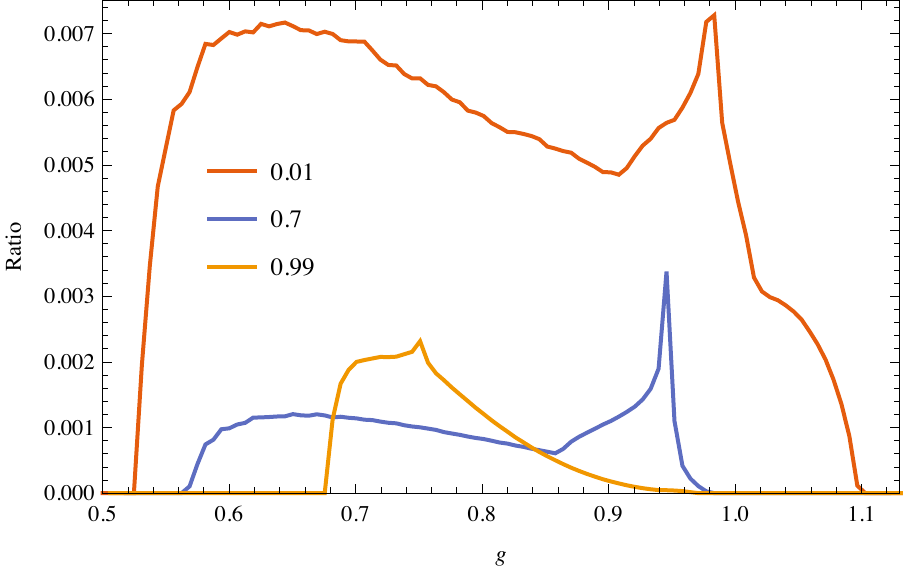}}
  \subfigure[\protect\label{fig:ratio2}]{ \includegraphics[width=7cm,height=5cm]{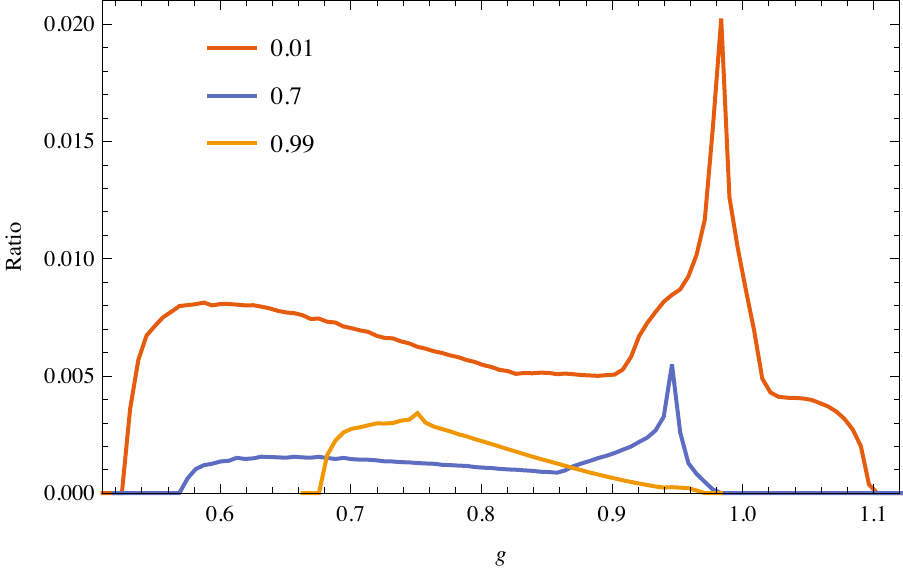}}
  \caption{Ratio between the flux of the second order image and the flux of all images. (a) Ratio evaluated for a limb darkened accretion disk. (b) Ratio evaluated for a limb brightened accretion disk.}\label{ratio}
\end{figure}

Now, we can quantify the error made by neglecting higher order images by calculating the ratio between the sum of the contributions of all higher order images and these profiles obtained by the lower order images only. For three representative inclinations, we find the plots shown in Fig. \ref{ratio}. From these figures it is clear that the contribution of higher order images reaches the maximum value for an edge-on accretion disk and decrease when the inclination increases, as also anticipated in Ref. \cite{fuerst}.

The relative contribution of higher order images remains below the percent level. The only exception comes from the edge-on limb-brightened case, where the peak exceeds $2\%$; in this case, the higher order images will build up a small bump at $g\approx0.9$. This detailed comparison between the contributions of higher and lower order images gives us some good material for the following discussion.

\section{Discussion and conclusions}

In this paper, we have proposed to use the strong deflection limit formalism to obtain accurate analytical results on the higher order images of the accretion disk. We have derived the magnification, redshift and emission angle functions in a fully analytical way. These functions enter the numerical integration where the radial and angular emissivity profiles can be specified according to our preferred disk model. We have shown that two different regimes exist for the line shape. For edge-on observers ($\mu_o<1/\sqrt{2}$, corresponding to $\vartheta_o<45^\circ$), the peak frequency is determined by a saddle point in the redshift function. For face-on observers ($\mu_o>1/\sqrt{2}$), the peak frequency is determined by the redshift at the ISCO. All these results do not depend on the model of accretion disk, because all the quantities only depend on the metric of space-time. Indeed, the ansatz on the intrinsic emissivity enters only in the numerical integration to determine the profile shape. So, our formalism can be applied to any physically motivated models of disks.

In the literature there are many works about the relativistic shapes of the Fe K$\alpha$ line. Some of them also try to address the role of higher order images of the accretion disk, but the results seem quite discrepant. Ref. \cite{beck} uses a full numerical approach to determine the shape of second-order line profile. Their profiles for the higher order contribution are dramatically different from those we have elaborated, with the presence of several peaks and more complex structures. On the other hand, we find a satisfying agreement with the results obtained by Bao et al. \cite{bao1994} where we can see that the structure of spectra for a quasi-equatorial observer is very similar to the one determined by us. We have been able to identify the origin of the lonely peak appearing in the contribution by higher order images, attributing it to the existence of a saddle point in the redshift function. The position of the peak is tracked by the redshift of this saddle point. The fact that we are proposing an approximation in the deflection angle (the strong deflection limit) so as to obtain simple analytical formulae has no effect on the overall structure, which is determined by the redshift function. The approximation intervenes in the calculation of the magnification, where it is accurate at the percent level, for the second and higher order images. Therefore, the structure of the contribution by higher order images appears relatively simple and we cannot imagine any physical origin for the multi-peak structures appearing in Ref. \cite{beck}, which might be the fruit of numerical errors. The existence of other independent calculations agreeing with our results \cite{bao1994} supports this conclusion.

These examples show how higher order images can be tricky to model in numerical codes for relativistic emission lines. Although they contribute by less than $1\%$ to the total flux in the line, they should not be overlooked at all, since at specific frequencies (namely, their peak frequency) they might give rise to localized bumps in the total line profile, as we can see in Fig. \ref{fig:ratio2}. For relatively bright sources such as Cygnus X-1, an accuracy of $1\%$ has already been reached \cite{Walton2016}. So, models trying to fit the observations should not neglect the contribution of higher order images.

In this work we have confined our investigation to the Schwarzschild black hole, in order to obtain simple and appealing analytical formulae. However, the strong deflection limit formalism has been extended to slowly rotating black holes \cite{Bozza2005,Bozza2006,Bozza2007}. This will allow us to widen our study to the Kerr case, where we hope to get similar analytical results that can be useful to get a deeper and thorough comprehension of the contributions of higher order images to the relativistic lines. The existence of wide extended caustics \cite{Bozza2008} hints at an even higher relevance of the contributions of higher order images.

\end{document}